# A Spatial-Temporal Analysis of Travel Time Gap and Inequality Between Public Transportation and Personal Vehicles[*]

Meiyu (Melrose) Pan (Corresponding Author), Christa Brelsford, Majbah Uddin
Oak Ridge National Laboratory, 1 Bethel Valley Road, Oak Ridge, TN 37830
Corresponding Author contact: panm@ornl.gov

**ABSTRACT**
The increased use of personal vehicles presents environmental challenges, prompting the exploration of public transportation as an affordable, eco-friendly alternative. However, obstacles like fixed schedules, limited routes, and extended travel times impede widespread adoption. This study investigates the temporal evolution of spatial inequality in the travel time gap between public transportation and personal vehicles, reflecting disparities across states and time periods. Analyzing Census Transportation Planning Program data for six northeastern states in 2010 and 2016 reveals no significant increase in the travel time gap, but notable growth in inequality in a few urban and disadvantaged communities. Comprehending these trends is vital for fostering equitable advancements in transportation infrastructure and enhancing public transportation competitiveness.

**INTRODUCTION**
The surge in personal vehicle use has sparked substantial environmental concerns, prompting the emergence of public transportation as an affordable and eco-friendly alternative. Despite its advantages, widespread adoption faces challenges like fixed schedules, limited routes, low population density, and commuter perceptions. Among these challenges, the notable hindrance to widespread public transportation utilization is the substantial travel time gap between public transportation and personal vehicles. In a comprehensive review, Redman et al. highlighted studies targeting various quality attributes, emphasizing speed as crucial for increasing ridership (Redman et al. 2013). For example, a New York City study revealed a 15-minute reduction in commuting time leading to a substantial 25% increase in rail service ridership (Liao et al. 2020).

   Measuring and comparing travel times between public transportation and personal vehicles is pivotal for assessing the efficiency and competitiveness of public transportation systems. A significant disparity in travel times between personal vehicles and public transportation could signify constrained mobility options, especially in rural areas, where accessibility to other regions may entail prolonged public transportation travel times, accentuating the challenges faced by individuals unable to afford a car or its associated expenses. While previous studies have delved into the travel time gap between public transportation and personal vehicles, fewer have explored the potential spatial inequalities of this gap. For instance, in urban areas with well-designed transit systems, certain regions may exhibit notably extended public transportation travel times to specific destinations (Dastgoshade, Hosseini-Nasab, and Mehrjerdi 2023). Averaging travel times may obscure variations in connectivity across subregions, underscoring the necessity of spatial inequality analysis for targeted improvements in public transportation connectivity.

---

[*] Notice: This manuscript has been authored by UT-Battelle, LLC, under contract DE-AC05-00OR22725 with the US Department of Energy (DOE). The US government retains and the publisher, by accepting the article for publication, acknowledges that the US government retains a nonexclusive, paid-up, irrevocable, worldwide license to publish or reproduce the published form of this manuscript, or allow others to do so, for US government purposes. DOE will provide public access to these results of federally sponsored research in accordance with the DOE Public Access Plan (https://www.energy.gov/doe-public-access-plan).

Numerous studies have examined public transportation service equity, proposing various methodologies and often focusing on specific regions or demographic groups (Hosein Mortazavi and Akbarzadeh 2017; Jin, Kong, and Sui 2019; Yeganeh et al. 2018). Different theories, such as utilitarianism, libertarianism, intuitionism, and Rawls' egalitarianism, have been employed to define equitable distribution of public transportation resources (Nahmias-Biran, Martens, and Shiftan 2017; Pereira, Schwanen, and Banister 2017).

Among these studies, one research gap exists in understanding the temporal changes in the travel time gap and the associated inequalities. While public transportation systems are generally improving, some areas may still experience uneven enhancements, leading to connectivity disparities (Zhang and Zhang 2021, 2022). Urbanization and inflation are identified as contributing factors to these inequalities resulting from improvements in the public transportation system (Lv et al. 2019; Mishra and Agarwal 2019). Analyzing temporal changes becomes imperative for developing comprehensive transit improvement strategies that consider evolving socio-economic and infrastructural landscapes.

In order to address these gaps in current research, this study aims to scrutinize the travel time gap between public transportation and personal vehicles, assess its spatial inequalities, and examine how these factors change over time. Leveraging the Census Transportation Planning Program (CTPP) dataset, we compare the travel time gap between 2010 and 2016 in six states in the northeastern U.S. In current literature, the concept of equity takes on diverse definitions regarding what is deemed "fair," typically associated with individual and/or household characteristics. However, this study primarily analyzes spatial inequality, centering on variations in transit competitiveness across regions, without delving specifically into individual and/or household considerations. Departing from the examination of individual or household-level travel time, this study aims to provide an overarching perspective by assessing multiple states. In addressing two key research questions:

(1) How does the travel time gap between public transportation and personal vehicles evolve over time and vary across states in the U.S.?
(2) Is there evidence of spatial inequality in the travel time gap, and how does its temporal trend unfold, with variations across states in the U.S.?

This study can contribute to the field by comprehensively assessing public transportation competitiveness across distinct time periods and scrutinizing the associated inequalities. Through this evaluation, the objective is to offer valuable insights into the evolving dynamics of public transportation competitiveness, supporting informed decision-making in urban planning and policy formulation.

## METHODOLOGY

**Data Source**

We leveraged data from the CTPP dataset, which procures tabulations of American Community Survey (ACS) 5-year (and historical Census decennial) data. Notably, the CTPP data stands out due to its inclusion of origin-destination flows from home to work at small geographies, differentiating it from ACS data. The dataset, designed to aid transportation analysts and planners, illuminates commuting patterns and modes of transportation.

It is important to note that this work serves as a preliminary study, focusing on a subset of states rather than aiming for a nationwide scope. This deliberate narrowing allows us to gain preliminary insights into the dynamics of public transportation competitiveness in specific regions. Our study specifically focuses in on six northeastern U.S. states: New York (NY), Maryland (MD), Pennsylvania (PA), Virginia (VA), New Jersey (NJ), and Massachusetts (MA). By concentrating our efforts on this regional subset, we aim to provide targeted and contextually relevant findings that can lay the groundwork for more extensive studies in the future.

We employed two series of CTPP data: 2010 (derived from the 2006-2010 American Community Survey) and 2016 (derived from the 2012-2016 American Community Survey). Each series contains the mean travel time for an origin-destination (OD) pair at the Census Tract level. The CTPP records data on five modes of transportation, each defined as follows (U.S. Census Bureau 2023):

- Drive alone: Individual occupancy of a car, truck, or van.
- Carpool: Involved a car, truck, or van with two or more individuals sharing the ride.
- Public transportation: Involved buses, trolley buses, streetcars, trolley cars, subways, elevated trains, railroads, or ferryboats.
- Taxi/Other: Involved taxicab, motorcycles or other unconventional methods.
- Bike/Walk: Traveled by biking or walking.

Our study specifically delves into analyzing travel times for two distinct modes of transportation: public transportation and drive alone, referred to as personal vehicles.

In order to mitigate bias, we excluded OD pairs that lacked data on travel times for public transportation or solo driving. This refinement yielded a dataset comprising 4,705 OD pairs, constituting 11.6% of the total. The majority of these exclusions were in regions with limited public transportation services, a reasonable decision allowing us to concentrate on comparisons in areas where public transportation services are available for meaningful evaluation.

Beyond state-level comparisons of travel time gap and inequality, our study also delved into the evaluation of disadvantaged communities (DAC). The data for identifying DAC originates from the Equitable Transportation Community (ETC) Explorer developed by the U.S. Department of Transportation (USDOT 2023). DACs are defined at the Census Tract level based on five components: Transportation Insecurity, Climate and Disaster Risk Burden, Environmental Burden, Health Vulnerability, and Social Vulnerability. The distribution of DACs and the study regions is visually represented in Figure 1.

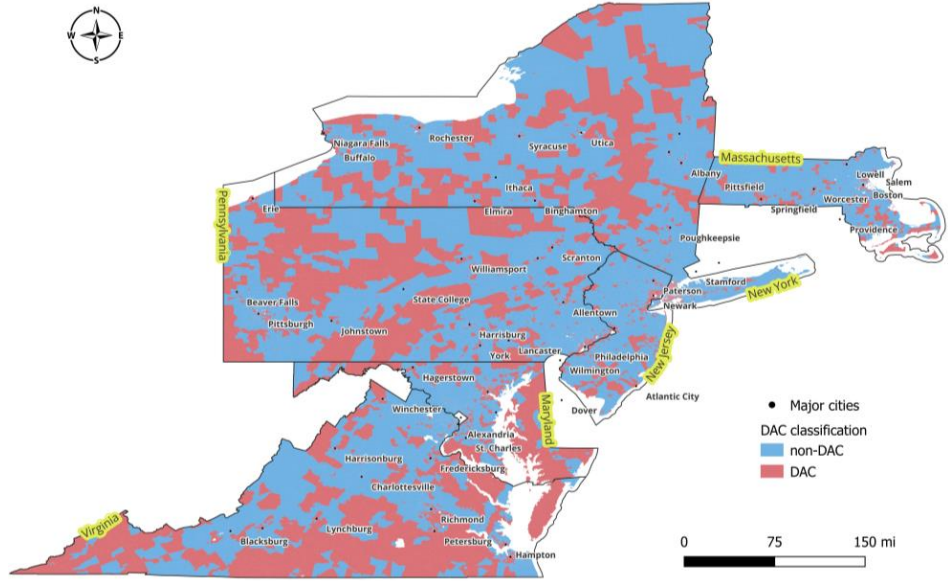

**Figure 1. DAC distribution and study states**

**Travel time gap and inequality Metrics**

The travel time gap, denoted as $g_{ab}$ for each OD pair, is established as a variable based on data at the Census Tract level. It is represented by the ratio of public transportation travel time over driving alone travel time. In order to capture travel time gap inequality within a broader region, such as a county or state, our goal is to identify a metric that effectively represents the heterogeneity.

While variance or standard deviation are commonly used as indicators of heterogeneity, we aim for a standardized measure for ease of comparison. Motivated by this consideration, we adopt the metric proposed by Pandey et al. (Pandey, Brelsford, and Seto 2022). The inequality of the travel time gap is denoted as $I$. This variable is constructed as a composite metric involving the mean ($\mu_g$) and standard deviation ($\sigma_g$) the travel time gap. This approach ensures that the metric is standardized by the average and theoretically bounded, facilitating straightforward comparisons. For the travel time gap, the base unit is the Census Tract to Census Tract OD pair. However, for inequality, the base unit is aggregated to the residence Census Tract. The two metrics are illustrated in Table 1. The theoretical range of this inequality metric spans from 0 (indicating the lowest inequality) to 1 (representing the highest inequality). As the standard deviation of the travel time gap increases within a region, the corresponding inequality metric also rises.

**Table 1. Illustration of the travel time gap and inequality metrics**

| Travel time gap | | Inequality | |
|---|---|---|---|
| Notation | Base unit | Formula | Base unit |
| $g$ | Census Tract to Census Tract OD pair | $I = \dfrac{\sigma_g}{\sqrt{\mu_g(1-\mu_g)}}$ | residence Census Tract |

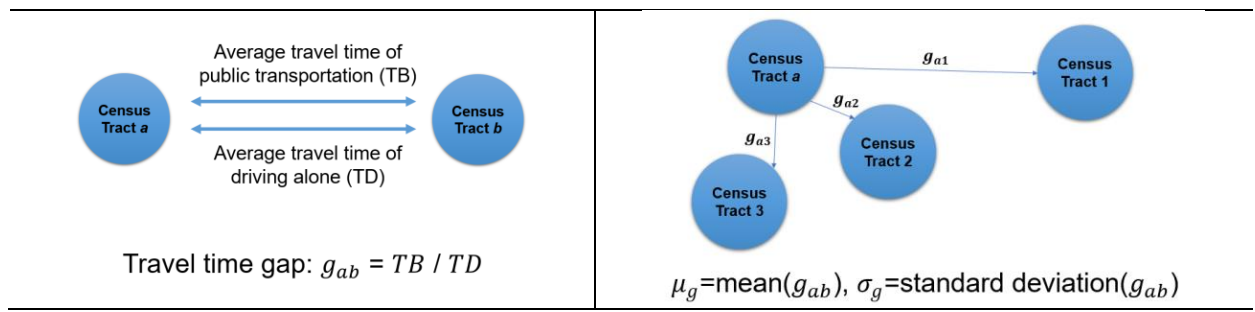

**Paired T-test Method Through Hypothesis Testing**

In this study, our focus is on testing the means of two key variables—travel time gap and inequality—between the years 2010 and 2016. Hypothesis testing, specifically the paired t-test, is employed for this purpose.

# RESULTS

The analysis of travel time between the two modes of transportation reveals a slight increase from 2010 to 2016 for both modes across all studied states. Despite this observed uptick, it's important to note that these changes are not statistically significant at 95% confidence level at the state level. The modest variations in travel time do not reach a level of significance that would allow us to confidently assert a meaningful difference. This suggests that, while there may be a slight temporal trend in travel time, it doesn't attain statistical significance when considering the variability at the state level.

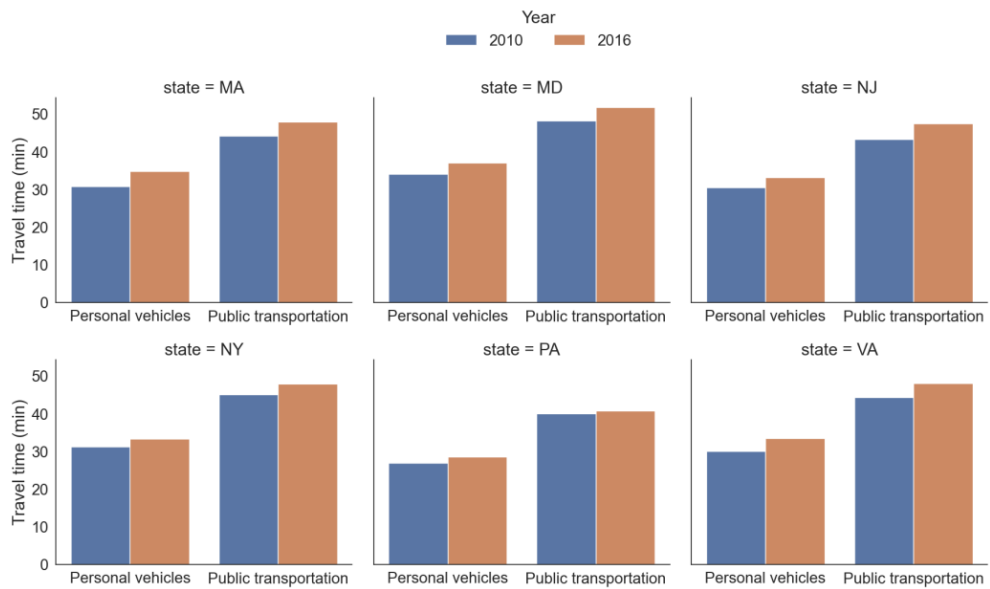

**Figure 2. Travel time comparison between personal vehicles and public transportation**

**Research Question 1: How does the travel time gap between public transportation and personal vehicles evolve over time and vary across states in the U.S.?**

The travel time gap across the six states is aggregated by residence Census Tract and presented Figure 3. The regions with available travel time data are predominantly situated in major cities and urban areas. This spatial concentration underscores that the analysis is focused on areas characterized by higher population density and urbanization, providing a targeted perspective on travel time gaps in these significant and interconnected locales.

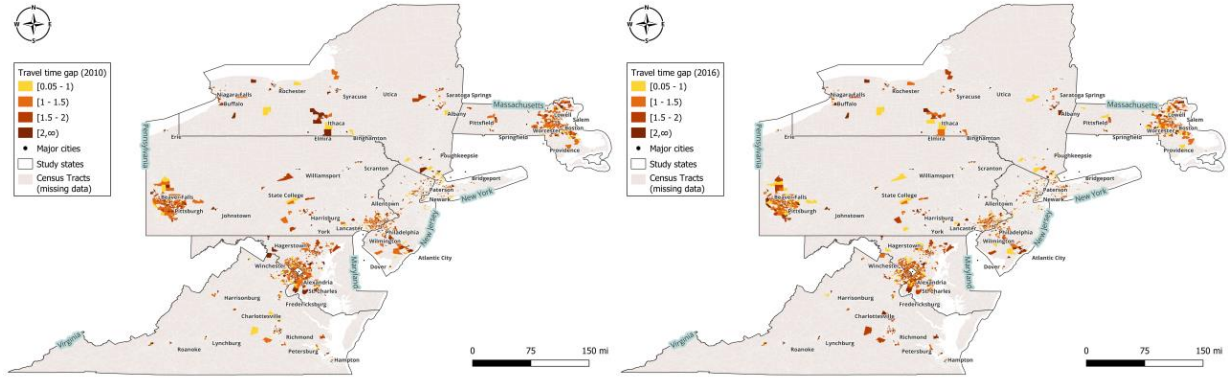

(a) Travel time gap in 2010  (b) Travel time gap in 2016

**Figure 3. Travel time gap temporal change**

Interestingly, from 2010 to 2016, both DAC and non-DAC regions underwent some change in travel time gap (Figure 4). While not statistically significant in all states, it indicates a noteworthy difference. Except for Pennsylvania, where DAC areas showed a decreasing trend (p-value=0.003), all other DAC areas demonstrated an increasing travel time gap trend. In contrast, all non-DAC areas exhibited a decreasing trend in travel time gap.

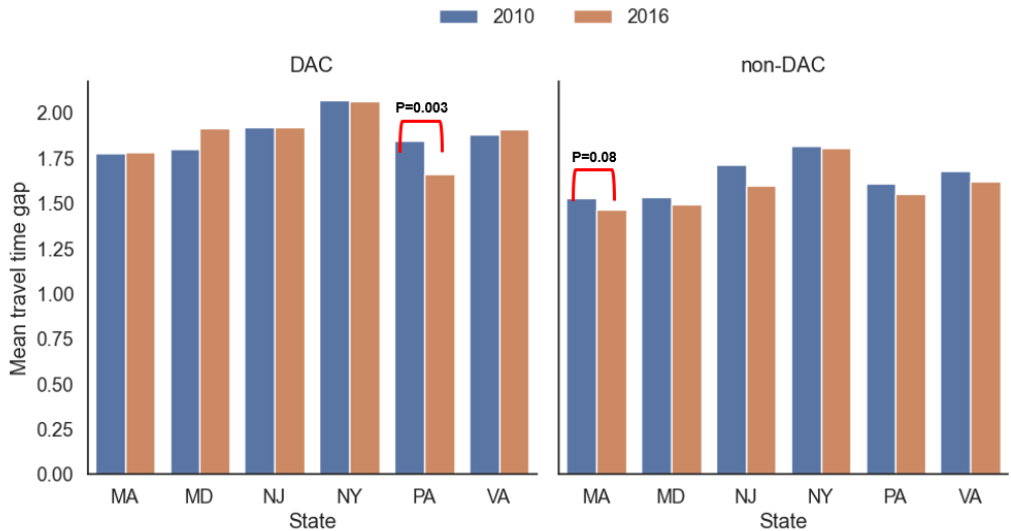

**Figure 4. Travel time gap between 2010 and 2016 in DAC and non-DAC areas**

When comparing the travel time gap between DAC and non-DAC areas in 2010, significant disparities emerge. DAC travel time gap is statistically higher than non-DAC travel time gap in New York ($p = 0.003$), Maryland ($p < 0.001$), Pennsylvania ($p < 0.001$), Massachusetts ($p < 0.001$), and New Jersey ($p = 0.04$). In terms of 2016, the DAC travel time gap remains significantly higher than the non-DAC travel time gap in all studied states, with p-values consistently lower than 0.001. This persistent significance underscores enduring disparities in travel time gaps.

**Research Question 2: Is there a presence of inequality in the travel time gap, and how does its temporal trend unfold?**

Figure 5 displays the distribution of mean and standard deviation of travel time gap inequality, along with the resulting inequality, across all six states. Although the mean travel time gap exhibits a decreasing trend, the standard deviation of the travel time gap is higher in 2016 compared to 2010, evident from the darker region in the figure. Consequently, there is a slight increase in inequality, as indicated by the shade area reaching slightly higher towards the "I=1" curve, representing the highest level of inequality. Figure 6 further illustrates the distribution of travel time gap inequality.

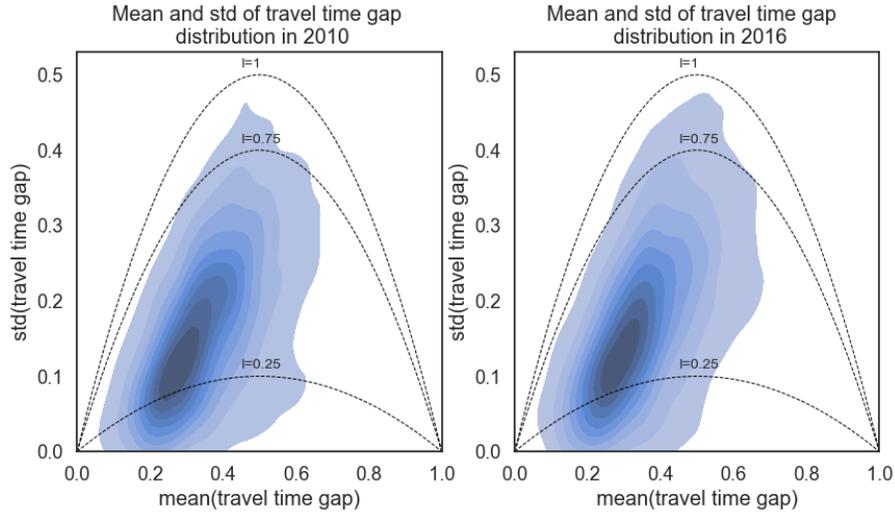

**Figure 5. Mean, standard deviation, and inequality**

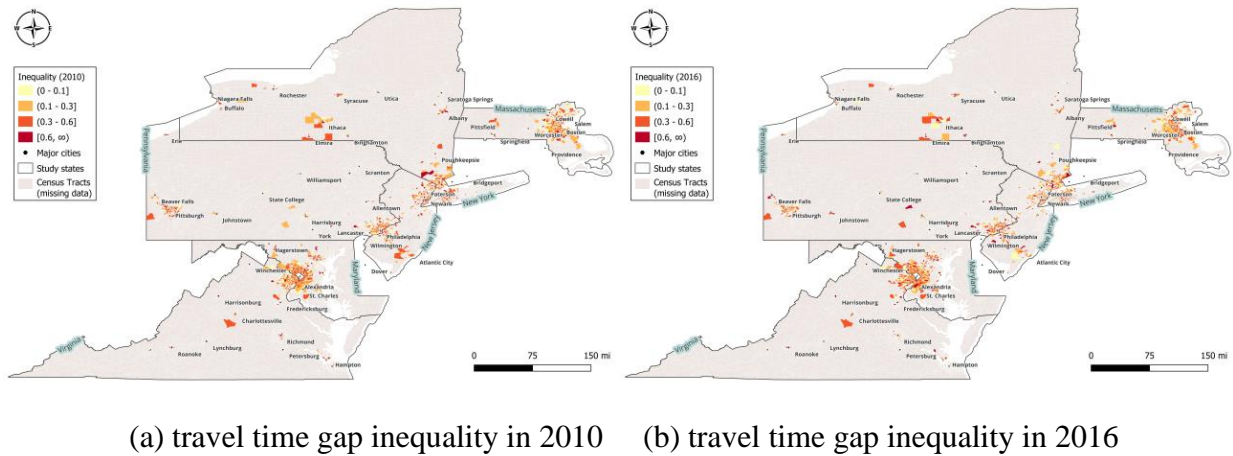

(a) travel time gap inequality in 2010    (b) travel time gap inequality in 2016

**Figure 6. Travel time inequality temporal change**

Non-DAC areas witness a significant uptick in inequality compared to DAC in Maryland ($p<0.001$), New York ($p =0.04$), and non-DAC in Massachusetts ($p =0.03$), as shown in Figure 7. In the DAC vs. non-DAC comparison, non-DAC areas show slightly higher inequality than DAC areas in both 2010 and 2016. Particularly, DAC inequality is significantly higher than non-DAC in 2010 for all studied states except New York and Maryland with *p*-values lower than 0.001. In 2016, DAC inequality remains significantly higher than non-DAC for all studied states except Pennsylvania and New Jersey with *p*-values lower than 0.001.

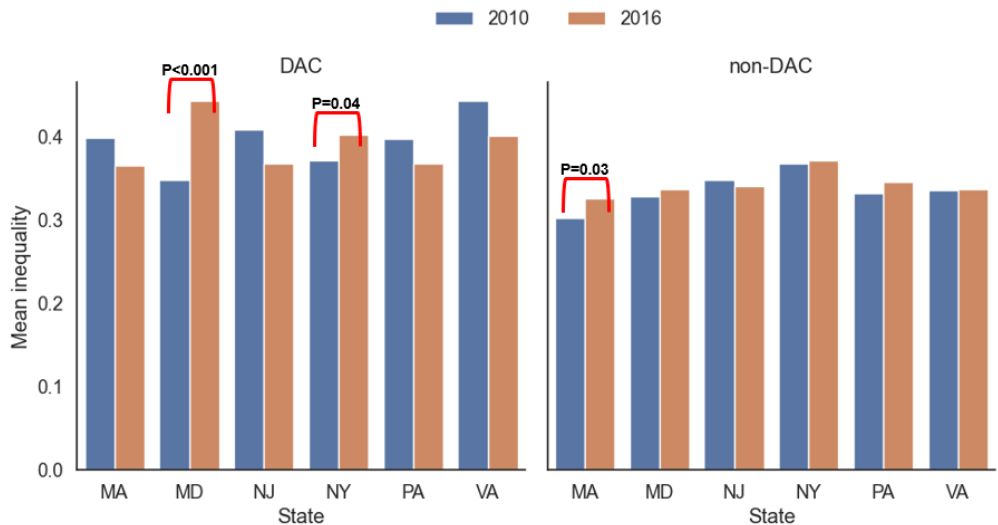

**Figure 7. Inequality between 2010 and 2016 in DAC and non-DAC areas**

When considering the trends of travel time gap and inequality together, a disparity emerges. In contrast to the decreasing trend observed in travel time gap, inequality demonstrates an increase. Specifically, DAC regions in Maryland and New York, despite no significant change in travel time gap, experience a notable increase in inequality. Furthermore, despite a significant decrease in travel time gap, non-DAC regions in Massachusetts simultaneously witness a significant increase in inequality from 2010 to 2017.

In summary, these findings suggest that while the travel time gap between public transportation and personal vehicles may not exhibit a significant increase over the years, there could be an uptick in inequality in certain regions. These results underscore the importance of considering more than just the mean, as spatial inequality may reflect uneven improvements in public transportation services over time. The observed phenomenon may be attributed to disproportionate enhancements in public transportation provisions across regions, especially in DAC areas. In DAC areas, there might be improvements in public transportation accessibility, but these enhancements may not extend to other potentially underserved regions. This multi-year analysis provides valuable insights for agencies to identify regions where future public transportation improvements may be warranted.

Moreover, both the travel time gap and inequality are higher in DAC areas than in non-DAC areas, suggesting that DAC areas experience lower connectivity via public transportation modes. This finding aligns with previous research indicating that people residing in DAC areas might have fewer economic opportunities, limited access to essential needs, and reduced ability for recreations (Oviedo and Sabogal 2020; Yousefzadeh Barri et al. 2021).

## CONCLUSIONS

This study examines temporal changes in travel time gap inequality between personal vehicles and public transportation, revealing a contrasting trend between travel time gap and inequality, as well as between DAC and non-DAC areas. While the travel time gap decreases, inequality increases, potentially due to disproportionate enhancements in public transportation, especially in DAC areas. Both the travel time gap and inequality are higher in DAC areas than in non-DAC areas, underscoring the lower connectivity in DAC areas and highlighting the disproportionate improvement in public transportation.

The paper contributes significantly to understanding temporal trends in public transportation competitiveness across regions, aiding agencies in identifying areas for targeted service improvements to boost ridership.

In Future work, we would like to extend this analysis to the entire U.S. Additionally, adopting more comprehensive equity metrics, connecting inequality to per capita measures, could enhance the spatial metric. Further exploration of factors influencing inequality can provide meaningful insights on improving overall equity in transportation services.